# Quantum-Material Josephson Junctions: Unconventional Barriers, Emerging Functionality

Kathryn A. Pitton[1,2], Michiel P. Dubbelman[1,2*], Trent M. Kyrk[1,2*], Houssam El Mrabet Haje[1,2*], Yaozu Tang[1,2], Roald J.H. van der Kolk[1,2], Yaroslav M. Blanter[1,2], Mazhar N Ali[1,2+]

[1]Department of Quantum Nanoscience, Faculty of Applied Sciences, Delft University of Technology, Lorentzweg 1, 2628 CJ, Delft, The Netherlands
[2]Kavli Institute of Nanoscience, Delft University of Technology, Lorentzweg 1, 2628 CJ, Delft, The Netherlands
*These authors contributed equally: Michiel P. Dubbelman, Trent M. Kyrk, Houssam El Mrabet Haje
[+]email: m.n.ali@tudelft.nl

**Abstract:**

Josephson Junctions translate quantum phase coherence into an electrical response and underpin superconducting sensors and quantum circuits. In conventional junctions, the barrier acts primarily as a passive weak link; however, when the barrier is a *quantum material* with its own internal degrees of freedom like magnetism, strong correlations, or switchable polarization, the Josephson effect becomes a sensitive probe of symmetry and many-body physics in the interlayer. Here we review progress in "quantum-material Josephson junctions," (QMJJ) focusing on three rapidly advancing barrier families: 1.) magnetic barriers, where exchange, noncollinearity, and spin-active scattering enable 0–π–φ ground states, singlet–triplet conversion, and nonreciprocal transport; 2.) correlated barriers, where proximity effects acquire many-body character and recent van der Waals Kagome Mott interlayers exhibit field-free Josephson diode behavior; and 3.) ferroelectric and multiferroic barriers, where nonvolatile polarization provides an internal control knob and can produce superconducting memory and memristive dynamics.

## Introduction:

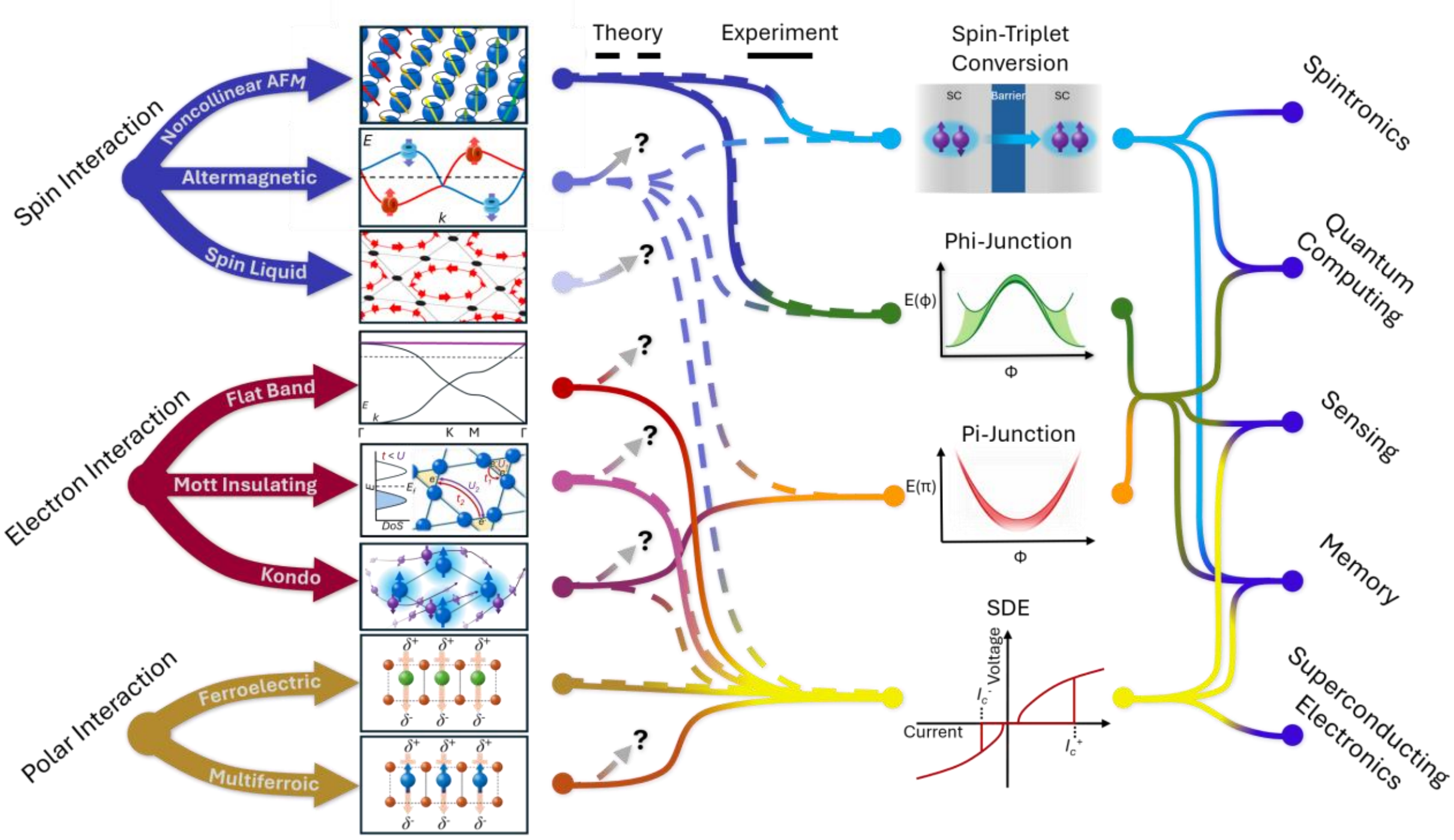

*Figure 1. Schematic of varying unconventional barrier classes and corresponding QMJJ phenomena. Dashed and solid lines highlight their respective theory and experimental paths. Arrows with the question marks suggest areas that have yet to be investigated, highlighting literature gaps.*

The Josephson junction (JJ)—two superconductors (SCs) coupled through a weak link—is a quintessential element of superconducting technology.[1] The DC Josephson effect creates a dissipationless charge current due to a phase difference between two SCs, forming the basis for JJ-based sensors such as Superconducting Quantum Interference Devices (SQUID). JJs also exhibit the AC Josephson effect, linking a constant voltage to an alternating supercurrent at a well-defined frequency that is directly proportional to the applied voltage.[2,3] This effect provides a universal phase voltage relationship linking quantum coherence to measurable electromagnetic radiation, enabling the generation and detection of microwave radiation, underpins voltage to frequency conversion, and serves as the operational basis for superconducting standards, high frequency detectors, and many superconducting Qbit architectures. Thus, the DC and AC JJs can be engineered for use in computing, sensing, metrology, and quantum information.[4]

While the macroscopic behavior of Josephson junctions is governed by phase coherence, the microscopic transport properties depend strongly on the nature of the barrier separating the superconductors. In JJs where the weak link between SCs is established through a non-superconducting barrier, the barrier properties govern the mechanism of charge transport: either tunneling in normal insulators or Andreev reflection in normal metals.[5,6] The insulating barrier can play an active role in determining the Josephson response, like in traditional magnetic insulating barriers where the spin structure can lead to nontrivial current–phase relations (CPR) which cannot be attributed solely to generic interfacial effects. By contrast in metal barriers, there is sufficient density of states at the Fermi level for Andreev reflection to dominate, where a Cooper pair is transmitted into the metallic barrier by the injection of an electron and the formation of a hole, followed by the electron-hole pair recombination at the other superconducting interface.[6] While these broad barrier classes exhibit specific CPRs, the material properties that can be used to modify the behavior of JJs are far richer and more complex, with topological and ferromagnetic materials among the most studied at this time.

However novel barrier materials, from unconventional magnets to multiferroics, offer new ways to manipulate Josephson behavior and enable emerging technologies. For example, complex magnetic barrier JJs can act as an active functional element that can promote spin-triplet superconductivity, realize π-junction behavior, and operate as a spin valve for superconducting spintronic applications.[7] Recently, altermagnets, characterized by a magnetic structure with zero net magnetization but with spin-split electronic bands arising from symmetry-protected, momentum-dependent spin polarization, are theorized to offer a platform for engineering spin-dependent superconducting transport without stray magnetic fields or strong uniform exchange-field pair-breaking effects typically associated with ferromagnetic junctions.[8] This combination enables novel avenues for manipulating the superconducting phase, generating unconventional CPRs, and realizing spin-selective or nonreciprocal Josephson effects within a magnetically compensated system.

Correlated barrier materials, in which strong electron–electron interactions renormalize the material's electronic properties, can directly reshape superconducting correlations through modulation of the pair amplitude, interface-pinned oscillations, and other effects, also resulting in non-traditional JJ behavior with emerging experiments in van der Waals correlated Kagome interlayers reporting field-free Josephson diode behavior. Ferroelectric (FE) materials have a macroscopic, remnant, and switchable internal electric polarization, which can couple to superconducting transport when incorporated as barriers in JJs. In ferroelectric JJs, this polarization enables tunneling electro resistance, a phenomenon widely exploited in normal state ferroelectric random-access memories and field-effect transistors, offering an additional nonvolatile control parameter for superconducting tunneling.[9] Finally, some materials exhibit further complexity by overlapping the broad domains above, such as in multiferroicity wherein two or more ferroic orders , such as ferromagnetism, ferroelectricity, or ferroelasticity, coexist within a single material. When integrated in JJs as barriers, such multiferroic systems provide additional internal degrees of freedom including unique phonon interactions, enabling coupled electric, magnetic, and structural control of the superconducting phase and current.[10]

In this review we briefly (1) highlight the theoretical and experimental work done on these unique barrier types, (2) describe their possible applications, and (3) draw attention to the challenges which could seed the realization of novel superconducting logic, quantum computing, and spintronic applications. At the core of these next-generation nanotechnologies are foundational quantum materials which serve as the platform from which to build. Figure 1 outlines a range of unconventional materials that can be used as novel barriers in designing QMJJs with effects that can be leveraged for real world applications.

**Magnetic Barriers**

Magnetic barrier materials influence superconductivity in JJs through the net exchange interaction experienced by a Cooper pair over the superconducting coherence length within the barrier (Schematic of various scenarios discussed here are showcased in Figure 2). Superconducting junctions incorporating ferromagnetic barriers (S/F/S) are the most studied type of superconductor/magnetic material/superconductor (S/M/S) JJs, as their non-zero magnetization provides a platform for investigating spin-dependent transport and superconducting proximity effects. In conventional ferromagnetic junctions, the collinear exchange field lifts spin degeneracy, leading to oscillatory and rapidly decaying superconducting pair correlations and the emergence of π-junction behavior and strong suppression of conventional spin-singlet superconductivity.[11,12]

In contrast, noncollinear magnets host spin textures in which the local moments are not aligned strictly parallel or antiparallel to one another, resulting in real-space spin textures like vortices, spin spirals, chiral textures, skyrmions, or noncoplanar spin configurations. These textures can arise from geometric frustration such as Kagome and triangular lattices, or from relativistic effects, such as Dzyaloshinskii–Moriya coupling.[13,14] Such systems often exhibit unconventional transport phenomena, including anomalous and topological Hall responses and emergent electrodynamics, even in the absence of net magnetization.[15] More recently, Kagome metals such as $AV_3Sb_5$ have attracted significant attention due to the coexistence of correlated charge order, superconductivity, and magnetic fluctuations, with evidence suggesting that noncollinear spin correlations contribute to their unusual electronic responses.[16]

Within JJs, spatially varying magnetization can transform superconducting correlations rather than simply suppress them. Experimentally, controllable magnetic noncollinearity has been achieved through engineered multilayer structures such as S/F/F′/F″/S junctions with misaligned magnetizations, synthetic antiferromagnets, or domain-wall configurations. In these systems, superconducting correlations can be viewed as spin-structured entities that evolve as they propagate through the barrier. While a uniform exchange field causes rapid dephasing of singlet pairs, spatially varying magnetization produces a phase accumulation across the barrier that can invert the sign of the first harmonic of the Josephson current, rotating the spin state and mixing singlet and triplet components. This can stabilize π-junction states and enable singlet–triplet conversion which can lead to long-range equal-spin triplet correlations.[12,17] These long-range triplet pairs can propagate through strongly spin-polarized materials, making multilayer ferromagnetic junctions a powerful platform for studying superconductivity in engineered spin textures. Additionally, when multiple spin-dependent channels compete, higher-order harmonics can become comparable in magnitude, producing ground-state phases intermediate between 0 and π (φ-junctions).[18,19]

Similar implementations have been recently investigated in antiferromagnets with varying Neel vectors.[20] Experimental signatures of such triplet-mediated supercurrents have been observed both in intrinsically noncollinear magnetic order and in junctions where noncollinearity is engineered at interfaces. Long-range Josephson supercurrents have been reported using the chiral noncollinear antiferromagnet $Mn_3Ge$ as the barrier, where triangular spin structures promote spin-triplet proximity effects without multilayer ferromagnetic engineering.[21,22] By contrast, long-range supercurrents observed using $CrO_2$, a collinear half metallic ferromagnet, are interpreted as arising from spin-active interfaces rather than intrinsic noncollinearity.[7] Similar spin-triplet behavior has been demonstrated in Co-based junctions with inserted weak ferromagnetic layers and in systems incorporating Ho layers with conical spin textures acting as "triplet injectors", both of which exhibit supercurrent that persists through thick ferromagnetic barriers.[23,24] Noncollinear magnetic textures also modify the magnetic-field dependence of the critical current $I_c(B)$, producing asymmetric or shifted Fraunhofer patterns that reflect the sensitivity of superconducting phase accumulation to the underlying spin structure. Magnetic noncollinearity therefore emerges as a powerful design parameter for phase batteries, programmable couplers, and other phase-sensitive SC elements.

Frustrated magnetic systems provide an additional route to noncollinear order. In these materials, lattice geometry or competing interactions prevent simultaneous minimization of all exchange energies, stabilizing complex spin configurations, strong fluctuations, and highly degenerate ground states.[25] Such systems can host unconventional phases including noncoplanar spin textures, quantum spin liquids, and chiral magnetic order, which are often accompanied by enhanced sensitivity to external perturbations such as magnetic fields, pressure, or strain.[26] However, they have been investigated little in the context of JJs. These symmetry-rich states provide a conceptual bridge to recently identified altermagnets, which exhibit spin-split electronic bands despite having zero net magnetization.

Altermagnets represent an emerging platform for Josephson physics because their symmetry-protected spin splitting can modify superconducting phase coherence without generating stray magnetic fields. Recent theoretical studies predict that altermagnetic barriers could induce nonreciprocal Josephson transport, enabling superconducting diode behavior even in the absence of external magnetic fields.[27] In addition to diode effects, altermagnetic barriers are predicted to enable singlet-triplet conversion, anomalous Josephson currents, and tunable 0-π transitions.[28] Their symmetry-driven spin textures may also stabilize finite-momentum Cooper pairing analogous to Fulde–Ferrell–Larkin–Ovchinnikov states, but arising from crystal symmetry rather than Zeeman fields.[29] Perhaps most intriguingly, altermagnetic JJs have been proposed as platforms for topological superconductivity and Majorana bound states in the absence of net magnetization. Recent theoretical work demonstrates that altermagnetic order can provide the effective spin polarization required to stabilize Majorana modes while avoiding many of the challenges associated with ferromagnetic or externally biased systems.[30] This combination of symmetry protection, spin selectivity, and magnetic field-free operation highlights altermagnets as a compelling material class for future superconducting quantum physics. However, this field is still emerging and experimental validation for many of these claims remains to be seen.

**Correlated Materials**

Another representative class of QMJJ barriers are strongly correlated materials. To date, JJs incorporating strongly correlated barriers (S/CB/S) have been explored primarily at the theoretical level, with only a limited number of experimental studies reported.[31–33] Early theoretical work modeled S/CB/S junctions as superconducting leads separated by a correlated barrier composed of several atomic layers, where correlations are introduced through on-site interactions tuned across a Mott metal–insulator transition.[31,32,34,35] In the weakly correlated metallic regime, the junction behaves similarly to an SNS system, exhibiting non-sinusoidal CPR that depends on the barrier thickness, with the maximum supercurrent occurring at a phase different from π/2. As the correlation strength increases and the barrier approaches a Mott insulating state, the CPR evolves toward a sinusoidal form characteristic of tunneling junctions. Near the metal–insulator transition, however, strong correlation effects can lead to interface-pinned oscillations of the pair amplitude and an enhancement of the characteristic voltage $I_c R_n$, which may exceed the Ambegaokar–Baratoff (AB) limit for specific barrier thicknesses. When the system enters the fully insulating regime, the CPR becomes sinusoidal, but the critical current is strongly suppressed relative to the AB prediction. Related DMFT results have also been obtained using Hubbard-model descriptions with Bethe-lattice self-consistency.[36]

Strong electronic correlations can also generate unconventional proximity effects in flat-band metal systems. While these systems have been largely absent from literature on JJs, some theoretical studies predict that they may exhibit enhanced superconducting proximity effects compared to conventional S/N junctions, where the de Gennes theory predicts zero proximity effect due to absence of band dispersion.[37] In such systems, the temperature dependence of the critical current can deviate strongly from conventional behavior and may even become non-monotonic. Some of these effects have been observed experimentally on twisted bilayer graphene (TBG) barrier Josephson junctions [38]. When the TBG barrier is tuned to the flat-band regime, the measured critical current is enhanced compared with expectations for a normal metallic barrier, indicating that flat-band and correlated electronic states can significantly modify Josephson transport.

Another system in which correlated barriers have been theoretically described is that of quantum dot (QD) JJs, where correlations arise through the Kondo effect. In this Kondo regime, the Josephson effect

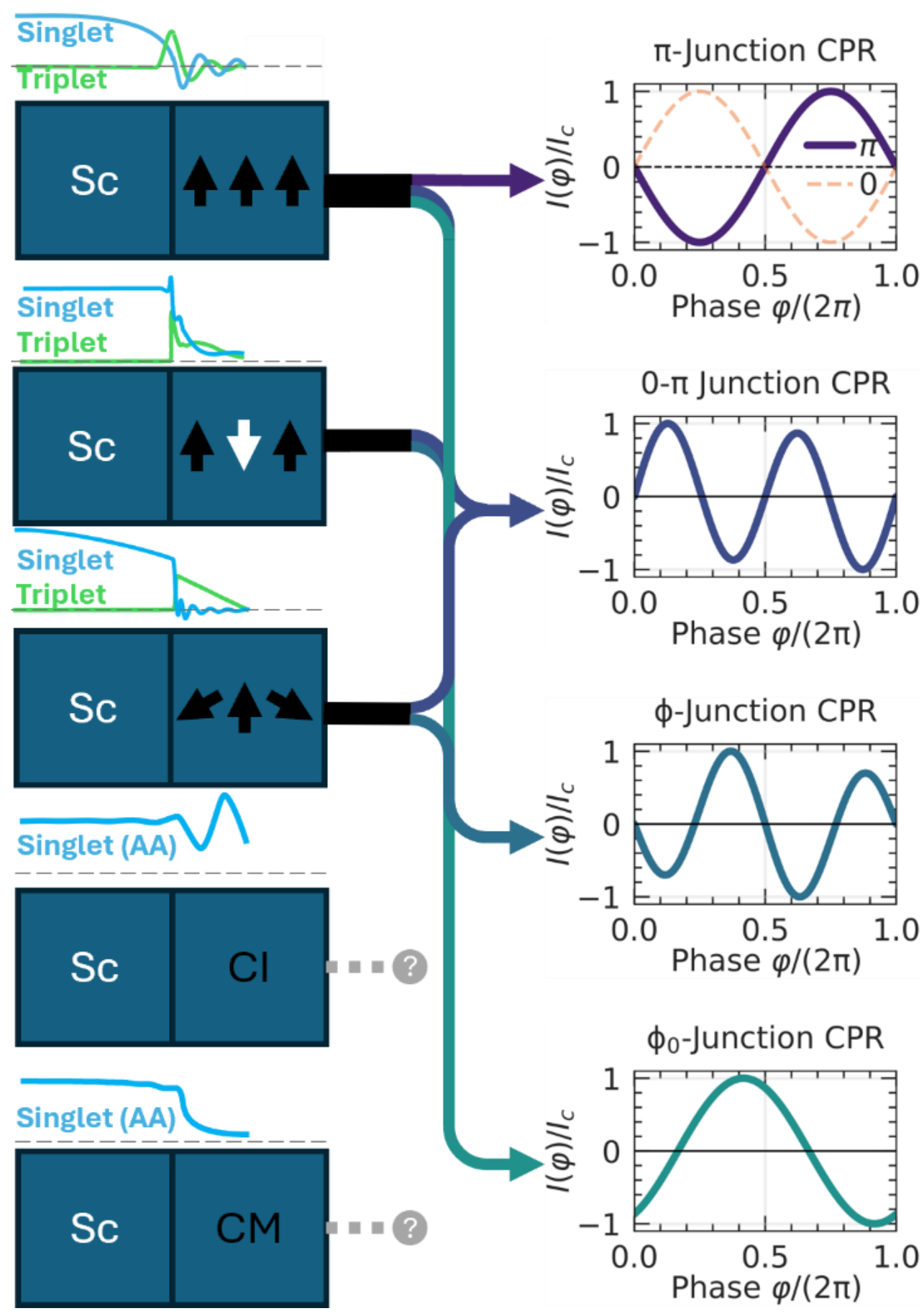

*Figure 2. Schematic of different QMJJs, with their spatial pair amplitude and possible CPR states. CI is correlated insulator and CM I correlated metal. For these two, the pair amplitude is approximated using the anomalous average (Freericks et al., 2001)*

can undergo a 0-π transition when the Kondo temperature becomes comparable to the superconducting critical temperature .[39,40] Outside the Kondo regime (Kondo temperature ($T_k$) << superconducting gap ($\Delta$)), the introduction of inversion-symmetry breaking through chirality provides an alternative route for correlations to modify the junction behavior. Here, chirality is treated as an intrinsic property of the quantum dot rather than arising from lead asymmetry, and no microscopic origin is specified, allowing for multiple possible experimental realizations. In this case, electronic correlations can not only induce a 0–π transition but also give rise to a field-free Josephson diode effect.[41] The magnitude of this effect depends both on the coupling strength between the superconducting leads and the QD, with a better coupling (relative correlation $U/\Delta$) leading to a bigger effect. Experimentally, the 0-π transition in the Kondo regime has been seen.[42,43] However, Josephson diode behavior in chirally coupled QD junctions has not yet been reported.

For correlated insulating barriers (S/CI/S), recent experiments on $NbSe_2/Nb_3X_8/NbSe_2$ junctions, where $Nb_3X_8$ (X = Cl, Br, I) is a breathing-mode Kagome lattice insulator, demonstrate strong modifications of the conventional Josephson effect.[33,41] These junctions exhibit a field-free Josephson diode effect, producing a pronounced asymmetry between the positive and negative critical currents ($I_{c+}$ and $I_{c-}$) even in the absence of an external magnetic field. The diode efficiency is largest for devices incorporating the most strongly correlated barrier and disappears for weakly correlated barriers. Further work has shown that this effect can be electrostatically tuned in $Nb_3Cl_8$ devices via gating.[33] Additionally, the temperature dependence of the critical current deviates strongly from the classical AB relation, with $I_c$significantly suppressed relative to conventional predictions.[41]

Superconductivity in Kagome correlated metals such as the $AV_3Sb_5$ family (A=K, Rb, Cs) has also attracted significant attention, with experiments suggesting intrinsic time-reversal symmetry breaking and the presence of chiral superconducting domains.[45–47]. However, because these studies probe the superconducting state of $AV_3Sb_5$ itself, it remains difficult to disentangle the microscopic origins of the observed phenomena; whether they arise from strong electronic correlations, unconventional magnetic order, or intrinsically chiral superconducting pairing. In contrast, there have been reports of employing non-stochiometric, non-superconducting $K_{1-x}V_3Sb_5$ as a barrier material in a JJ.[48] These junctions exhibit a strongly anisotropic magnetic field response: an in-plane field shows a somewhat traditional Fraunhofer pattern, while an out-of-plane field induces rapid oscillations with a critical current minimum near zero field.[48] This behavior is consistent with the emergence of anisotropic internal magnetic field, together with edge currents in the $K_{1-x}V_3Sb_5$.

These results indicate that electronic correlations can strongly modify the CPR.[34] Although no direct experimental probe of CPR in these correlated barrier systems has yet been reported, indirect signatures such as pronounced near-zero-field minima in the critical current suggest the presence of higher harmonics or phase shifted Josephson coupling. When inversion symmetry breaking is introduced, theoretical models further predict the emergence of nonreciprocal Josephson transport and additional higher-order contributions to the CPR. Much like prior, the temperature dependence of the JJ also deviates from conventional AB behavior.[49] The magnitude and nature of these deviations depend sensitively on the strength of electronic correlations, which has not yet been systematically explored experimentally. Consequently, Josephson junctions incorporating correlated metals such as $Ni_3In$ or $YbRh_2Si_2$, as well as Mott insulators including $VO_2$ or 1T-$TaS_2$, represent promising platforms for isolating correlation-driven effects and uncovering new Josephson phenomena. In general, the AB model and conventional scattering descriptions of the Josephson effect become insufficient in junctions with correlated barriers, as their underlying assumption of non-interacting electrons no longer applies. In this regime, the barrier cannot be viewed as a merely unconventional tunneling region but instead acts as an interacting quantum medium that imprints symmetry, topology, and many body effects onto the superconducting response. Their use in JJs opens a pathway toward devices in which phase, spin, and directionality are intrinsically linked, rather than externally imposed. As experimental control over correlation strength, symmetry breaking, and interface quality continues to improve, correlated-barrier JJs are poised to become a versatile platform for discovering and exploiting new superconducting phenomena.

**Ferroelectric Barriers**

Recent advances in ferroelectric growth and fabrication, together with the emergence of exfoliable van der Waals ferroelectrics/multiferroics, have enabled new opportunities to explore phenomena arising at ferroelectric-superconductor interfaces, for example, by using the former as barriers in JJs.[50–52] We are now just at the beginning of this research direction, and below we present the early work.

Several phenomena can arise in Josephson junctions incorporating ferroelectric barriers. Building on the well-established theory of ferroelectric tunnel junctions (FTJs), recent theoretical work predicts that in asymmetric JJs the critical current can be modulated by the ferroelectric polarization, producing polarization-dependent supercurrents, as illustrated in Figure 3a.[53,54] This asymmetry can be engineered through junction design, for example by introducing insulating layers with different thicknesses. Such polarization-controlled supercurrents enable gate-tunable Josephson transport, which has already been demonstrated experimentally.[51,55] At ferroelectric–superconductor interfaces, polarization-induced surface charges can also modify superconducting properties such as the critical temperature and critical current, providing a nonvolatile control mechanism that has been exploited in cryogenic memory devices.[55,56] Furthermore, feedback between the Josephson current and the polarization state can generate nonlinear dynamics. Early studies predicted oscillations between polarization states, while more recent theoretical work based on the resistively and capacitively shunted junction (RCSJ) model suggests memristive behavior in ferroelectric Josephson junctions, opening the possibility of superconducting ferroelectric memristors.[57]

Experimentally, the first demonstrations of ferroelectric barriers into JJs have already been reported (Figure 3b,c).[55,58–61] Although challenges remain, these results open the door to hybrid ferroelectric–superconducting devices such as superconducting diodes, field-effect transistors, and cryogenic nonvolatile memories. The versatility of these systems also provides multiple control knobs, including electrostatic gating and THz excitation, which have already been demonstrated experimentally. Additional approaches may soon be explored, such as optical control of ferroelectric order or magnetic-field tuning of ferroelectric properties in multiferroic Josephson junctions.

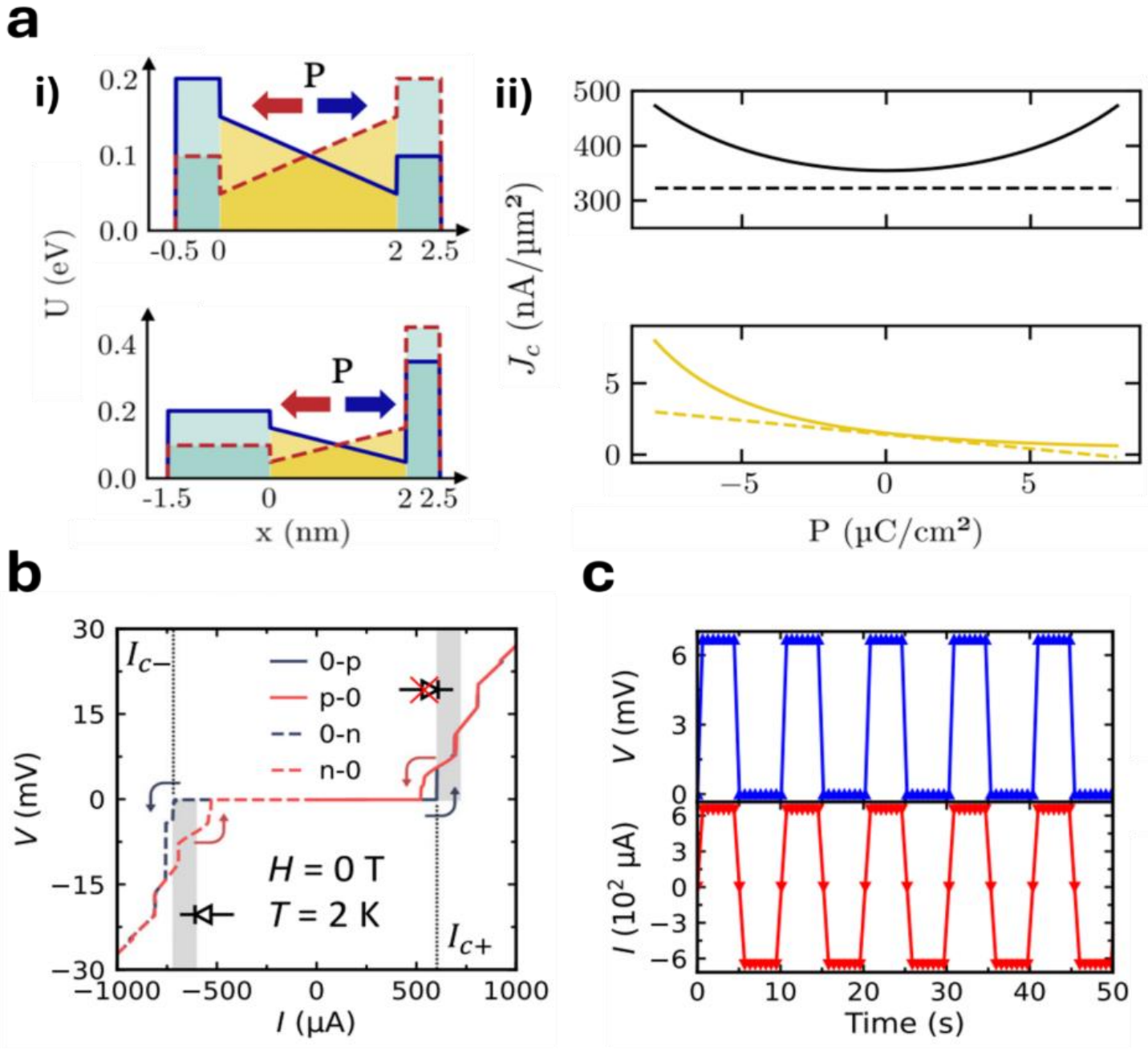

Figure 3: (a) Polarization-controlled supercurrent in a ferroelectric Josephson junction. (i) Schematic potential profiles for symmetric (top) and asymmetric (bottom) dielectric–ferroelectric–dielectric barriers. Blue (solid) and red (dashed) curves represent opposite polarization directions. (ii) Calculated critical current density as a function of ferroelectric polarization for symmetric and asymmetric barriers (solid: numerical calculation; dashed: analytical approximation).[54] (b) $V$–$I$ characteristics of the $NiI_2$ junction. Current sweep directions (0–p, p–0, 0–n, n–0) correspond to $0 \rightarrow +1000\mu$A, $+1000 \rightarrow 0\mu$A, $0 \rightarrow -1000\mu$A, and $-1000 \rightarrow 0\mu$A. The critical currents $I_c^{+600\mu A}$and $I_c^{-=718\mu A}$ are defined by the first switching events in the 0–p and 0–n curves. The shaded region marks the diode operating range. (c) Demonstration of supercurrent rectification $I_{\mathrm{bias}} = \pm 650\mu$A. [55,58–61]

**Conclusion:**

This review summarized how unconventional magnetic order, electronic correlations, and polarization can modify Josephson transport beyond conventional tunnel-junction behavior. Across ferromagnetic, noncollinear, altermagnetic, correlated, and ferroelectric barriers, the superconducting current–phase relation reflects the underlying spin, correlation, and polarization landscape of the barrier material. Recent experiments demonstrating phase shifts, higher-harmonic contributions, and nonreciprocal Josephson effects show that these junctions can access regimes of superconducting transport beyond those described by standard proximity-effect models. Continued advances in materials growth, interface control, and phase-sensitive measurements will be essential for disentangling these effects and for establishing Josephson junctions with complex magnetic or correlated barriers as controlled platforms for exploring unconventional superconductivity and related quantum phenomena.

References

(1) Tinkham, M. *Introduction to Superconductivity: Second Edition*; Courier Corporation, 2004.
(2) Josephson, B. D. Possible New Effects in Superconductive Tunnelling. *Physics Letters* **1962**, *1* (7), 251–253. https://doi.org/10.1016/0031-9163(62)91369-0.
(3) Shapiro, S. Josephson Currents in Superconducting Tunneling: The Effect of Microwaves and Other Observations. *Phys. Rev. Lett.* **1963**, *11* (2), 80–82. https://doi.org/10.1103/PhysRevLett.11.80.
(4) Likharev, K. K. Superconducting Weak Links. *Rev. Mod. Phys.* **1979**, *51* (1), 101–159. https://doi.org/10.1103/RevModPhys.51.101.
(5) Ambegaokar, V.; Baratoff, A. Tunneling Between Superconductors. *Phys. Rev. Lett.* **1963**, *10* (11), 486–489. https://doi.org/10.1103/PhysRevLett.10.486.
(6) Andreev, A. F. Thermal Conductivity of the Intermediate State of Supercondcutors. *Zh. Eksperim. i Teor. Fiz.* **1964**, *Vol: 46*.
(7) Keizer, R. S.; Goennenwein, S. T. B.; Klapwijk, T. M.; Miao, G.; Xiao, G.; Gupta, A. A Spin Triplet Supercurrent through the Half-Metallic Ferromagnet CrO2. *Nature* **2006**, *439* (7078), 825–827. https://doi.org/10.1038/nature04499.
(8) Šmejkal, L.; Sinova, J.; Jungwirth, T. Emerging Research Landscape of Altermagnetism. *Phys. Rev. X* **2022**, *12* (4), 040501. https://doi.org/10.1103/PhysRevX.12.040501.
(9) Tsymbal, E. Y.; Kohlstedt, H. Tunneling Across a Ferroelectric. *Science* **2006**, *313* (5784), 181–183. https://doi.org/10.1126/science.1126230.
(10) Garcia, V.; Fusil, S.; Bouzehouane, K.; Enouz-Vedrenne, S.; Mathur, N. D.; Barthélémy, A.; Bibes, M. Giant Tunnel Electroresistance for Non-Destructive Readout of Ferroelectric States. *Nature* **2009**, *460* (7251), 81–84. https://doi.org/10.1038/nature08128.
(11) Birge, N. O. Spin-Triplet Supercurrents in Josephson Junctions Containing Strong Ferromagnetic Materials. *Phil. Trans. R. Soc. A.* **2018**, *376* (2125), 20150150. https://doi.org/10.1098/rsta.2015.0150.
(12) Buzdin, A. I. Proximity Effects in Superconductor-Ferromagnet Heterostructures. *Rev. Mod. Phys.* **2005**, *77* (3), 935–976. https://doi.org/10.1103/RevModPhys.77.935.
(13) Dzyaloshinsky, I. A Thermodynamic Theory of “Weak” Ferromagnetism of Antiferromagnetics. *Journal of Physics and Chemistry of Solids* **1958**, *4* (4), 241–255. https://doi.org/10.1016/0022-3697(58)90076-3.
(14) Moriya, T. Anisotropic Superexchange Interaction and Weak Ferromagnetism. *Phys. Rev.* **1960**, *120* (1), 91–98. https://doi.org/10.1103/PhysRev.120.91.
(15) Nagaosa, N.; Sinova, J.; Onoda, S.; MacDonald, A. H.; Ong, N. P. Anomalous Hall Effect. *Rev. Mod. Phys.* **2010**, *82* (2), 1539–1592. https://doi.org/10.1103/RevModPhys.82.1539.
(16) Ortiz, B. R.; Teicher, S. M. L.; Hu, Y.; Zuo, J. L.; Sarte, P. M.; Schueller, E. C.; Abeykoon, A. M. M.; Krogstad, M. J.; Rosenkranz, S.; Osborn, R.; Seshadri, R.; Balents, L.; He, J.; Wilson, S. D. CsV3Sb5: A Z:2 Topological Kagome Metal with a Superconducting Ground State. *Phys. Rev. Lett.* **2020**, *125* (24), 247002. https://doi.org/10.1103/PhysRevLett.125.247002.
(17) Bergeret, F. S.; Volkov, A. F.; Efetov, K. B. Odd Triplet Superconductivity and Related Phenomena in Superconductor-Ferromagnet Structures. *Rev. Mod. Phys.* **2005**, *77* (4), 1321–1373. https://doi.org/10.1103/RevModPhys.77.1321.
(18) Konschelle, F.; Buzdin, A. Magnetic Moment Manipulation by a Josephson Current. *Phys. Rev. Lett.* **2009**, *102* (1), 017001. https://doi.org/10.1103/PhysRevLett.102.017001.
(19) Bergeret, F. S.; Tokatly, I. V. Theory of Diffusive Φ0 Josephson Junctions in the Presence of Spin-Orbit Coupling. *EPL* **2015**, *110* (5), 57005. https://doi.org/10.1209/0295-5075/110/57005.
(20) Linder, J.; Robinson, J. W. A. Superconducting Spintronics. *Nature Phys* **2015**, *11* (4), 307–315. https://doi.org/10.1038/nphys3242.
(21) Sukhanov, A. S.; Singh, S.; Caron, L.; Hansen, T.; Hoser, A.; Kumar, V.; Borrmann, H.; Fitch, A.; Devi, P.; Manna, K.; Felser, C.; Inosov, D. S. Gradual Pressure-Induced Change in the Magnetic Structure of the Non-Collinear Antiferromagnet Mn$_3$Ge. *Phys. Rev. B* **2018**, *97* (21), 214402. https://doi.org/10.1103/PhysRevB.97.214402.

(22) Jeon, K.-R.; Hazra, B. K.; Cho, K.; Chakraborty, A.; Jeon, J.-C.; Han, H.; Meyerheim, H. L.; Kontos, T.; Parkin, S. S. P. Long-Range Supercurrents through a Chiral Non-Collinear Antiferromagnet in Lateral Josephson Junctions. *Nat. Mater.* **2021**, *20* (10), 1358–1363. https://doi.org/10.1038/s41563-021-01061-9.

(23) Khaire, T. S.; Khasawneh, M. A.; Pratt, W. P.; Birge, N. O. Observation of Spin-Triplet Superconductivity in Co-Based Josephson Junctions. *Phys. Rev. Lett.* **2010**, *104* (13), 137002. https://doi.org/10.1103/PhysRevLett.104.137002.

(24) Robinson, J. W. A.; Witt, J. D. S.; Blamire, M. G. Controlled Injection of Spin-Triplet Supercurrents into a Strong Ferromagnet. *Science* **2010**, *329* (5987), 59–61. https://doi.org/10.1126/science.1189246.

(25) Ramirez, A. P. Strongly Geometrically Frustrated Magnets. *Annu. Rev. Mater. Sci.* **1994**, *24* (1), 453–480. https://doi.org/10.1146/annurev.ms.24.080194.002321.

(26) Moessner, R.; Ramirez, A. P. Geometrical Frustration. *Physics Today* **2006**, *59* (2), 24–29. https://doi.org/10.1063/1.2186278.

(27) Banerjee, S.; Scheurer, M. S. Altermagnetic Superconducting Diode Effect. *Phys. Rev. B* **2024**, *110* (2), 024503. https://doi.org/10.1103/PhysRevB.110.024503.

(28) Ouassou, J. A.; Brataas, A.; Linder, J. Dc Josephson Effect in Altermagnets. *Phys. Rev. Lett.* **2023**, *131* (7), 076003. https://doi.org/10.1103/PhysRevLett.131.076003.

(29) Zhang, S.-B.; Hu, L.-H.; Neupert, T. Finite-Momentum Cooper Pairing in Proximitized Altermagnets. *Nat Commun* **2024**, *15* (1), 1801. https://doi.org/10.1038/s41467-024-45951-3.

(30) Ghorashi, S. A. A.; Hughes, T. L.; Cano, J. Altermagnetic Routes to Majorana Modes in Zero Net Magnetization. *Phys. Rev. Lett.* **2024**, *133* (10), 106601. https://doi.org/10.1103/PhysRevLett.133.106601.

(31) Freericks, J. K.; Nikolić, B. K.; Miller, P. Temperature Dependence of Superconductor-Correlated Metal–Superconductor Josephson Junctions. *Appl. Phys. Lett.* **2003**, *82* (6), 970–972. https://doi.org/10.1063/1.1543236.

(32) Tahvildar-Zadeh, A. N.; Freericks, J. K.; Nikolić, B. K. Thouless Energy as a Unifying Concept for Josephson Junctions Tuned through the Mott Metal-Insulator Transition. *Phys. Rev. B* **2006**, *73* (18), 184515. https://doi.org/10.1103/PhysRevB.73.184515.

(33) Wu, H.; Wang, Y.; Xu, Y.; Sivakumar, P. K.; Pasco, C.; Filippozzi, U.; Parkin, S. S. P.; Zeng, Y.-J.; McQueen, T.; Ali, M. N. The Field-Free Josephson Diode in a van Der Waals Heterostructure. *Nature* **2022**, *604* (7907), 653–656. https://doi.org/10.1038/s41586-022-04504-8.

(34) Freericks, J. K.; Nikolić, B. K.; Miller, P. Tuning a Josephson Junction through a Quantum Critical Point. *Phys. Rev. B* **2001**, *64* (5), 054511. https://doi.org/10.1103/PhysRevB.64.054511.

(35) Miller, P.; Freericks, J. K. Microscopic Self-Consistent Theory of Josephson Junctions Including Dynamical Electron Correlations. *J. Phys.: Condens. Matter* **2001**, *13* (13), 3187. https://doi.org/10.1088/0953-8984/13/13/326.

(36) Rolih, D.; Žitko, R. Strongly Correlated Josephson Junction: Proximity Effect in the Single-Layer Hubbard Model. arXiv February 16, 2026. https://doi.org/10.48550/arXiv.2602.14796.

(37) Li, Z. C. F.; Deng, Y.; Chen, S. A.; Efetov, D. K.; Law, K. T. Flat Band Josephson Junctions with Quantum Metric. *Phys. Rev. Res.* **2025**, *7* (2), 023273. https://doi.org/10.1103/PhysRevResearch.7.023273.

(38) Díez-Carlón, A.; Díez-Mérida, J.; Rout, P.; Sedov, D.; Virtanen, P.; Banerjee, S.; Penttilä, R. P. S.; Altpeter, P.; Watanabe, K.; Taniguchi, T.; Yang, S.-Y.; Law, K. T.; Heikkilä, T. T.; Törmä, P.; Scheurer, M. S.; Efetov, D. K. Probing the Flat-Band Limit of the Superconducting Proximity Effect in Twisted Bilayer Graphene Josephson Junctions. *Phys. Rev. X* **2025**, *15* (4), 041033. https://doi.org/10.1103/ccb4-tqxq.

(39) Choi, M.-S.; Lee, M.; Kang, K.; Belzig, W. Kondo Effect and Josephson Current through a Quantum Dot between Two Superconductors. *Phys. Rev. B* **2004**, *70* (2), 020502. https://doi.org/10.1103/PhysRevB.70.020502.

(40) Lee, M.; López, R.; Xu, H. Q.; Platero, G. Proposal for Detection of the 0' and Pi'Phases in Quantum-Dot Josephson Junctions. *Phys. Rev. Lett.* **2022**, *129* (20), 207701. https://doi.org/10.1103/PhysRevLett.129.207701.
(41) Dubbelman, M. P.; Wu, H.; Aretz, J.; Wang, Y.; Pasco, C. M.; Zhao, Y.; Kyrk, T. M.; Yang, J.; Xu, X.; McQueen, T. M.; Roesner, M.; Ali, M. N. Driving the Field-Free Josephson Diode Effect Using Kagome Mott Insulator Barriers. arXiv December 18, 2025. https://doi.org/10.48550/arXiv.2512.17099.
(42) Buitelaar, M. R.; Nussbaumer, T.; Schönenberger, C. Quantum Dot in the Kondo Regime Coupled to Superconductors. *Phys. Rev. Lett.* **2002**, *89* (25), 256801. https://doi.org/10.1103/PhysRevLett.89.256801.
(43) Kanai, Y.; Deacon, R. S.; Oiwa, A.; Yoshida, K.; Shibata, K.; Hirakawa, K.; Tarucha, S. Electrical Control of Kondo Effect and Superconducting Transport in a Side-Gated InAs Quantum Dot Josephson Junction. *Phys. Rev. B* **2010**, *82* (5), 054512. https://doi.org/10.1103/PhysRevB.82.054512.
(44) Wu, S.; Ren, Z.-H.; Yang, L.; Wang, M.-Y.; Zhang, X.-P.; Fan, X.-Y.; Zhang, H.-C.; Li, X.; Wang, G.; Wang, C.; Li, C.; Wang, Z.-W.; Li, C.-Z.; Liao, Z.-M.; Yao, Y.-G. Efficiency-Tunable Field-Free Josephson Diode Effect in Nb3Cl8 Based van Der Waals Junctions. *Nano Lett.* **2025**, *25* (51), 17619–17627. https://doi.org/10.1021/acs.nanolett.5c04336.
(45) Blom, T. J.; Rog, M.; Altena, M.; Salinas, A. C.; Wilson, S. D.; Li, C.; Lahabi, K. Emergent Network of Josephson Junctions in a Kagome Superconductor. arXiv October 10, 2025. https://doi.org/10.48550/arXiv.2510.09777.
(46) Le, T.; Xu, Z.; Liu, J.; Zhan, R.; Wang, Z.; Lin, X. Thermomodulated Intrinsic Josephson Effect in Kagome CsV3S5. *Phys. Rev. Lett.* **2025**, *135* (9), 096002. https://doi.org/10.1103/yqth-sfm8.
(47) Wu, H.; Haje, H. el M.; Dubbelman, M.; Ortiz, B. R.; Wilson, S. D.; Ali, M. N.; Wang, Y. Supercurrent Interference and Its Transfer in a Kagome Superconductor. arXiv October 12, 2025. https://doi.org/10.48550/arXiv.2510.10543.
(48) Wang, Y.; Yang, S.-Y.; Sivakumar, P. K.; Ortiz, B. R.; Teicher, S. M. L.; Wu, H.; Srivastava, A. K.; Garg, C.; Liu, D.; Parkin, S. S. P.; Toberer, E. S.; McQueen, T.; Wilson, S. D.; Ali, M. N. Anisotropic Proximity–Induced Superconductivity and Edge Supercurrent in Kagome Metal, K1−xV3Sb5. *Science Advances* **2023**, *9* (28), eadg7269. https://doi.org/10.1126/sciadv.adg7269.
(49) Chirolli, L.; Greco, A.; Crippa, A.; Strambini, E.; Cuoco, M.; Amico, L.; Giazotto, F. Diode Effect in the Fraunhofer Pattern of Disordered Planar Josephson Junctions. *Commun Phys* **2025**, *8* (1), 483. https://doi.org/10.1038/s42005-025-02364-y.
(50) Wang, C.; You, L.; Cobden, D.; Wang, J. Towards Two-Dimensional van Der Waals Ferroelectrics. *Nat. Mater.* **2023**, *22* (5), 542–552. https://doi.org/10.1038/s41563-022-01422-y.
(51) Qi, L.; Ruan, S.; Zeng, Y.-J. Review on Recent Developments in 2D Ferroelectrics: Theories and Applications. *Advanced Materials* **2021**, *33* (13), 2005098. https://doi.org/10.1002/adma.202005098.
(52) Xu, Y. *Ferroelectric Materials and Their Applications*, 1st ed.; Elsevier, 1991.
(53) Ktitorov, S. A.; Trepakov, V. A.; Ioffe, A. F.; Jastrabik, L.; Soukup, L. Josephson Effect in a Junction with a Ferroelectric. *Ferroelectrics* **1994**, *157* (1), 387–392. https://doi.org/10.1080/00150199408229537.
(54) Tang, Y.; Ali, M.; Bauer, G.; Blanter, Y. Polarization-Controlled Supercurrent in Ferroelectric Josephson Junction. *Phys. Rev. B* **2026**. https://doi.org/10.1103/nvbk-27dx.
(55) Suleiman, M.; Sarott, M. F.; Trassin, M.; Badarne, M.; Ivry, Y. Nonvolatile Voltage-Tunable Ferroelectric-Superconducting Quantum Interference Memory Devices. *Applied Physics Letters* **2021**, *119* (11), 112601. https://doi.org/10.1063/5.0061160.
(56) Golod, T.; Iovan, A.; Krasnov, V. M. Single Abrikosov Vortices as Quantized Information Bits. *Nat Commun* **2015**, *6* (1), 8628. https://doi.org/10.1038/ncomms9628.
(57) Badarne, M. A.; Dalla Torre, E. G.; Ivry, Y. Hybrid Superconducting-Ferroelectric Quantum Memristor. *Phys. Rev. Res.* **2025**, *7* (4), 043234. https://doi.org/10.1103/4746-w9cm.

(58) Paghi, A.; Borgongino, L.; Strambini, E.; Simoni, G. D.; Sorba, L.; Giazotto, F. The Ferroelectric Superconducting Field Effect Transistor. arXiv July 7, 2025. https://doi.org/10.48550/arXiv.2507.04773.

(59) Yang, H.-Y.; Cuozzo, J. J.; Bokka, A. J.; Qiu, G.; Eckberg, C.; Lyu, Y.; Huyan, S.; Chu, C.-W.; Watanabe, K.; Taniguchi, T.; Wang, K. L. Field-Resilient Supercurrent Diode in a Multiferroic Josephson Junction. *Nat Commun* **2025**, *16* (1), 9287. https://doi.org/10.1038/s41467-025-63698-3.

(60) Ding, S.; Yao, J.; Bi, Z.; Tran, Q.; Liu, B.; Li, Q.; Trolier-McKinstry, S.; Jackson, T. N.; Liu, Y. Indirect Tunneling Enabled Spontaneous Time-Reversal Symmetry Breaking and Josephson Diode Effect in TiN/Al2O3/Hf0.8Zr0.2O2/Nb Tunnel Junctions. arXiv December 12, 2025. https://doi.org/10.48550/arXiv.2504.16987.

(61) Karthikeyan, M.; Chanda, G.; Kaewmaraya, T.; Moontragoon, P.; Sun, G.; Li, Z.; Watcharapasorn, A. Systematic Modulation of Superconducting Gap Dynamics in YBCO|BNT|YBCO Josephson Junctions through THz Field Interaction and BNT Ferroelectric Barrier. Research Square October 1, 2025. https://doi.org/10.21203/rs.3.rs-7638991/v1.

**Acknowledgements**: The authors wish to thank Yaojia Wang and Heng Wu for feedback. The authors would also like to thank the Kavli Foundation for their support through the Kavli Institute Innovation Award, The Kavli Institute of Nanoscience Delft, The NWO Talent Programme VIDI financed by the NWO VI.Vidi.223.089, and the “Materials for the Quantum Age” (QuMat, registration number 024.005.006) which is part of the Gravitation program financed by the Dutch Ministry of Education, Culture and Science.

**Author Contributions**: KAP, MPD, TMK, HMH, YT, and RJHK designed, wrote, formatted and edited the manuscript. YB and MNA reviewed the manuscript. All authors contributed to the preparation of this manuscript.